# Disciplinary authenticity: Enriching the reforms of introductory physics courses for life-science students

Jessica Watkins[i], Janet E. Coffey[ii], Edward F. Redish[i], Todd J. Cooke[iii]

[i] Department of Physics, University of Maryland, College Park, Maryland 20742
[ii] Department of Curriculum and Instruction, University of Maryland, College Park, Maryland 20742
[iii] Department of Cell Biology & Molecular Genetics, U/ of Maryland, College Park, Maryland 20742

**Abstract.** Educators and policy-makers have advocated for reform of undergraduate biology education, calling for greater integration of mathematics and physics in the biology curriculum. While these calls reflect the increasingly interdisciplinary nature of biology research, crossing disciplinary boundaries in the classroom carries epistemological challenges for both instructors and students. In this paper we expand on the construct of authenticity to better describe and understand disciplinary practices, in particular to examine those used in physics and biology courses. We then apply these ideas to examine an introductory biology course that incorporates physics and mathematics. We characterize the uses of interdisciplinary tools in this biology course and contrast them with the typical uses of these tools in physics courses. Finally, we examine student responses to the use of mathematics and physics in this course, to better understand the challenges and consequences of using interdisciplinary tools in introductory courses. We link these results to the reform initiatives of introductory physics courses for life-science students.

## INTRODUCTION

In 2009 the National Academies published a report advocating for a new vision of the biological discipline to meet the economic and social challenges of the 21$^{st}$ century . In particular, the report envisioned a new approach to research in biology that draws from many scientific disciplines and collaborations across organizations:

> *The essence of New Biology is integration — re-integration of the many subdisciplines of biology, and the integration into biology of physicists, chemists, computer scientists, engineers, and mathematicians to create a research community with the capacity to tackle a broad range of scientific and societal problems. (p. vii)*

As part of this call for reform, the National Academies recommended further changes in how scientists are educated and trained, particularly at the undergraduate level. They advocated for more interdisciplinary courses, greater integration of biology in introductory physics (likewise, more physics and mathematics in introductory biology), and pedagogies centered on solving complex, real-world scientific problems that require interdisciplinary tools. These recommendations followed other national reports evaluating and re-envisioning science education [1-5], most notably the 2003 National Research Council report, *Bio2010: Transforming Undergraduate Education for Future Research Biologists,* and the 2009 report by AAMC-HHMI, *Scientific Foundations for Future Physicians.* These reports have the potential to dramatically impact physics education, specifically the introductory courses for life-science majors. For example, the AAMC-HHMI report recommended a shift in requirements for pre-medical students from the completion of specific courses to demonstration of various competencies. Notably, they called for students to demonstrate specific mathematics and physics proficiencies, many of which are not currently taught in undergraduate biology and pre-medical programs [6]. In response to the calls for change, the physics education community has started conversations about transforming introductory physics courses for life-science majors (IPLS courses) [7]. Conferences such as the 2009 Conference on Physics in Undergraduate Quantitative Life Science Education, and special sessions at meetings of the American Physical Society and the American Association of Physics Teachers have provided platforms for curriculum developers and instructors to share materials and approaches that incorporate more connections to biology.

In addition to the instructional challenges posed for IPLS courses, the physics education research community also faces new challenges. Transforming these courses will not only require adding new content, but bring together different ways of thinking about, building, and interacting with scientific knowledge. While there have been great advances in understanding student epistemologies in physics [8-12] and in developing new pedagogical approaches to address these ideas [13], the reforms of IPLS courses necessitate new, complementary research directions at the boundary of physics and biology education. In science education, the research that has examined disciplinary practices has primarily treated science as a homogeneous field, ignoring the differences between the disciplines, such as in [14], [15]. While there are commonalities across the sciences, there also exist important differences even among subdisciplines that can impact how these subjects are taught. For example, Mayr argues that evolutionary biology, as a historical science, focuses more on building and testing historical narratives than on the experimental methodology prevalent in physical sciences or functional biology [16]. Therefore, alongside interdisciplinary reforms, there needs to be research efforts to understand how to characterize disciplinary differences in science, particularly in classroom activities. For IPLS courses, understanding the authentic practices of biology, physics, and mathematics will enable instructors to make disciplinary differences explicit and help students navigate and make connections between the disciplines. Ultimately, as these courses are reformed, we need to examine how students perform in, interact with, and respond to these new interdisciplinary curricular environments, building on and comparing to previous research in physics.

In this paper, we offer a theoretical lens for considering differing disciplinary practices across the sciences. The goal of this work is to enhance the development of and research on interdisciplinary science courses. For IPLS course reforms, we can use this lens to guide the needed investigations into the different approaches physics and biology courses take when using mathematics and physics. For example, understanding which physics ideas biology courses use and how they use them is critical in developing IPLS courses that bridge these disciplines. Here we examine an introductory biology course that incorporates physics and mathematics to understand organismal biology. Our analysis serves two purposes: (1) to illustrate how our theoretical approach can be used to characterize disciplinary practices in the science classroom and (2) to provide greater insight on the intersection of the disciplinary practices of biology and physics. Importantly, we also examine student responses to the use of mathematics and physics in this course. Characterizing interdisciplinary contexts and investigating student responses to them will allow us to better understand the challenges and consequences of using interdisciplinary tools in introductory courses. To make sense of how and when to cross disciplinary boundaries in the physics classroom, we need productive ways of thinking and organizing our understanding about the scientific disciplines. In this section, we build on previous work on disciplinary authenticity [17], [18], which will help us consider similarities and differences among physics, mathematics, and the biological subdisciplines. While Brown, Collins, and Duguid introduce authentic activities as the "ordinary practices of the culture," (p. 34) we expand on their description to better identify and characterize authenticity in scientific disciplines. We define authentic activities in science as those that use tools — such as concepts, equations, or physical tools — in ways and for purposes that reflect how the disciplines build, organize, and assess knowledge about the world. With this characterization, disciplinary authenticity resides in the activities of the participants, as situated in the broader research community and as reflected in the course context.

We start our discussion by examining a physics exam problem, which allows us to unpack how students are typically asked to use the tools of physics in IPLS courses and discuss how these uses reflect the disciplinary practices of physics. We then use this as a touchstone example as we expand the construct of disciplinary authenticity to account for both the different tools and the different uses of the same tools across the science disciplines.

### Analyzing an introductory physics problem for life-science majors

In Figure 1 we present an exam problem taken from an introductory physics course for life-science majors.[1] Answering this question correctly requires thinking about the physical situation — understanding the motion of both the cheetah and the pronghorn. While part (A) of the problem is a one-step calculation,[2] part (B) involves decomposing the problem into smaller steps, each of which relates to the physical situation, e.g. figuring out the distance the cheetah can cover while accelerating. Once each step is figured out, the different pieces then need to be synthesized to fully understand how they relate to one another and to the question at hand.

Part (B) of this exam problem also asks students to translate between their interpretation of the physical situation and the mathematical formalisms. This translation requires that the equations mean more than just symbols and their definitions—the "*d*" is not just some distance, but how far the cheetah can run. Furthermore, in each of the different pieces of the solution, students have to alternate between the physics and math to see how the structure of the physical situation organizes and constrains how the kinematic equations can be applied. For example, the same equation can be used in multiple ways depending on whether the animal is accelerating or at a constant velocity. Although this problem requires reasoning with the equations, the mathematics is used in service of understanding the physical situation and calculating specific parameters to characterize it.

---

[1] This problem was posed in a course that was reformed to help students to learn to think scientifically. The instructors explicitly focused on the epistemology of physics, asking students to participate in ways that reflect the activities of physics. See Ref. [13]
[2] This one-step calculation is meant to prime the students away from using the acceleration formula for distance ($\frac{1}{2}at^2$) in the next part and to use instead the more conceptual formula for accelerated distance ($v_{av}\,\Delta t$).

---

The pronghorn antelope in the Western Great Plains of the US is one of the fastest animals on the planet. But it has outlived all its predators and now runs where none pursues. Let's imagine a cheetah-like predator in the period tens of thousands of years ago when the pronghorn evolved its speed.

A cheetah is one of the fastest animals, but it can only maintain its high speed for a short time. The pronghorn can continue to run at a steady pace of 80 km/hr for a long time. Here are some of the parameters of the cheetah's motion.

- Max speed 120 km/hr
- Can accelerate from 0 to 120 km/hr in 3 seconds
- Can maintain max speed (sprint) for about 30 seconds.
- After its initial high-speed sprint, it quickly drops to a steady pace of ~70 km/hr.

(A) During the time it is accelerating, what is the cheetah's average acceleration, $<a>$, and its average speed, $<v>$? Show your work in the space below and put your answers in the box at the right.

(B) Suppose the cheetah comes across a herd of antelope running at a steady pace of 80 km/hr. As soon as they see the cheetah, they wheel and run directly away from the cheetah. How far from the cheetah do the pronghorns have to be if they are to be safe? Explain your reasoning.

Fig. 1. An exam problem from a reformed IPLS course.

Finally, this problem also involves several simplifications and assumptions. For example, students must assume that the pronghorn do not need time to accelerate and that the cheetah's deceleration from 120 km/hr to 70 km/hr occurs on a negligible time scale. By not including the relevant information, this problem sends cues to the students that these assumptions are valid, at least for the purposes of this exam. The problem restricts the scope of the cheetah's chase of the pronghorn to a handful of motions and parameters. This allows students to focus on how the distance between an accelerating object and an object at constant speed changes over time.

This exam problem, which requires multistep reasoning, translating between the physical situation and mathematics, and making simplifying assumptions, reflects specific aspects of the epistemology of physics. It asks students to engage in activities that mirror how physicists create new knowledge via problem-solving. The results of these activities are also grounded in the physics discipline. Students need to use the tools of physics, including mathematical formalisms, to obtain a numerical solution for the class of initial configurations needed to reach a given final outcome. Furthermore, solving this problem allows for deeper physical understanding of how distance varies between objects that are at changing or constant velocities. This problem was a part of an exam with additional kinematics problems, but the other problems involved toy cars or pulling boxes to examine the relationships between force, acceleration, velocity, and position.

### Using authenticity to understand disciplinary differences

In this one exam problem, students are asked to use mathematical and physical tools in particular ways and for specific ends that are authentic to the physics discipline. These activities are authentic in the way that Brown, Collins, and Duguid [17] describe as reflecting the practices of a culture. In the culture of physics, specific ways of thinking, building knowledge, and communicating that knowledge are emphasized and valued. While these activities and evaluations have been and are continuing to be negotiated by members of the physics community, there exists a set of practices and values that are shared within the physics discipline and make it distinct even from the other sciences.

Brown, et al. use the construct of authenticity to indict the distinctions between school activities and those of practitioners. They argue that many of the activities that students engage in while in school "would not make sense or be endorsed by the cultures" of the discipline to which they are ascribed. Citing the situated nature of knowledge, Brown, et al. advocate for thinking of knowledge and concepts not as abstract entities, but as conceptual tools whose meanings are inextricably linked to the context and culture in which they are used. They draw attention to differences in how tools are used, as a function of the culture and activities in which they were developed. "The occasions and conditions for use arise directly out of the context of activities of each community that uses the tool, framed by the way members of that community see the world." (p. 33) Therefore, to be able to appropriately use the tools of a discipline, students need to learn these conceptual tools as a part of the communities and cultures that employ them, by engaging in the authentic activities of practitioners.

Researchers in science education have built on this idea of disciplinary authenticity [18] to examine scientific inquiry in research and in classrooms [19-23]. Chinn and Malhotra [19] infer and compare the cognitive processes in scientific research and typical classroom inquiry tasks, finding that the reasoning processes evoked are qualitatively different. The authors determined authenticity based on whether a task would evoke a given reasoning process, such as generating research questions. They developed these criteria by drawing from literature in philosophy and history of science, but primarily treat science as a unitary discipline. Other conceptualizations of disciplinary authenticity may use different criteria, yet still do not parse between the different science disciplines and subdisciplines [20], [22], [23]. Lee and Songer [21] take a more narrow approach at characterizing authenticity when looking at tasks in meteorology. They examined student explanations for features that were characteristic in professional meteorologists' explanations, such as consideration of multiple meteorological entities. Their criteria for determining authenticity was less focused on the specific practices of meteorology, but rather what knowledge students used to create explanations.

Applying the idea of authenticity to the calls for interdisciplinary science curricula for biology students, it becomes critical to understand the activities of biologists, physicists, and other scientists—and the similarities and differences between them. Brown et al. touch upon the disciplinary specificity of authentic activities and tool use, citing that physicists and engineers use mathematical formulae differently. However, their framework needs additional tools for unpacking disciplinary differences in authenticity. How do we determine which practices are authentic to a discipline? How do we then describe the authenticity of these practices? Furthermore, how do we make sense of similarities and differences among disciplinary practices? We build on their work to expand and specify what we take as authenticity, grounding this construct in the epistemology of different science disciplines.

*What tools to use*

While disciplinary and subdisciplinary boundaries in science may seem artificial and artifacts of the structural constraints of academia, there are substantive differences that allow for the parsing of science into different (yet overlapping) communities and cultures. Examining what tools different disciplines and subdisciplines use to understand phenomena, build new knowledge, and solve problems is a first step at unpacking what it means to authentically engage in the different disciplinary practices. To investigate the feet of a gecko, physicists may use the conceptual tools of van der Waals forces or capillarity to understand how the gecko can climb smooth surfaces, while an evolutionary biologist may focus on the history of the lamellae on the toes, comparing the modern-day gecko with the fossils of its prehistoric ancestors. The physicist and biologist may be studying similar phenomena, but use different tools, which will ultimately enable them to understand different, yet complementary, aspects of the gecko.

Of course, there is a great deal of overlap in the conceptual tools used across the scientific disciplines. For example, many disciplines can use the same principle, such as the Second Law of Thermodynamics. Furthermore, as discussed in the introduction, biology as a discipline is being called to use more of the tools of mathematics and physics. In the gecko example, understanding van der Waals forces and capillarity helps provide biologists with a deeper understanding of the evolutionary affordances and constraints that physics imposes. Often the goals of integrating the disciplines is to allow for the sharing the tools of one discipline to help broaden the scope of another. In fact, research in gecko adhesion is an excellent example of how interdisciplinary research teams can collaborate to make significant advances in science [24].

Much of the policy reports and conversations about the transformations of introductory physics courses have centered on what tools from physics would be most helpful and relevant to biology. However, just as Brown et al. cited that carpenters and cabinet-makers can use chisels differently; scientists may use the same conceptual tools differently, depending on the goals and purposes for their use.

*For what ends are these tools used*

In our gecko example, the biologist and physicist are not just using different tools to examine the gecko, but there also may be differences in their goals of their analyses. From a physics and engineering perspective, focus on the mechanism of gecko adhesion is paramount to understand the fundamental forces causing this specific phenomenon and to engineer new nanomaterials that mimic the gecko's stickiness [25]. From an evolutionary perspective, biologists may be striving to understand how geckos function in their natural environment, which entails not just narrowly understanding the material properties, but also how these properties fit in with knowledge about the entire organism, its evolutionary history, and the environmental constraints of its surroundings [26]. Therefore, a materials scientist and evolutionary biologist may use the same conceptual tool of van der Waals force, but for different purposes. This underscores that authenticity lies in how the tool is used and for what purpose, and not in the tool itself – in this case, the concept of van der Waals force. It would not suffice to consider tools independent from the activity and problem the tool is being used to address [17], [27].

Applying these ideas to the classroom, looking at the broader contexts in which a problem is situated will therefore be critical when thinking about authenticity. However, these problem contexts must be characterized not just by the features of the problem, but also what activities are required and, importantly, for what ends. For example, a final exam in a calculus class for biology students asked the problem shown in Figure 2. Its purpose is to ask students to set up an integral relating density to total amount. While the problem context appears biology-related, the activities that students engage in to solve this problem are only related to obtaining the mathematical expression, not to applying the expression to make sense of the effects of the population distribution of fish, for example. For these different purposes how one needs to think about and use this equation changes. As noted in [28], scientists interested in the fish population would likely attend to the units and be dismayed that the numbers represent different objects throughout the equation: "1" in the numerator represents length, while the "1" in the denominator is an area. However, if the purpose is solely to manipulate the equation and obtain an integral as part of practicing mathematical techniques, the mismatch in units may not a relevant issue. If, however, this sort of activity is designed to prepare students for the use of math in science classes, and what results is that students learn to ignore units and attend only to numerical value, the units become pedagogically if not mathematically significant.

---

The population density of trout in a stream is

$$r(x) = \frac{1+x}{x^2+1}$$

where *r* is measured in trout per mile and *x* is measured in miles. *x* runs from 0 to 10.
(a) Write an expression for the total number of trout in the stream. Do not compute it.
(b) ...

---

Fig. 2. A final exam problem from an introductory calculus course.

In efforts to make IPLS courses more connected to needs of biology students, consideration of authenticity will require unpacking the different ends for which the disciplines can use the same or similar tools. This examination will help clarify how scientific disciplines and subdisciplines can support the goals and purposes of each other. Understanding how physics and mathematics can be used *in the service of* biological understanding will point instructors toward productive connections to biologically-authentic activities. Furthermore, understanding the ends to which a tool is used will guide how to use it.

*How to use these tools*

To understand how disciplines use tools differently, we can first examine how they may attend to different characteristics of a tool. With an equation, for example, a physicist may attend only to the leading terms to understand the physical implications, while a mathematician, in search of greater precision, may be more likely to attend to the higher order terms. We can describe how physicists and mathematicians use an equation differently partly by what characteristics they attend to in their use in given contexts.[3]

Furthermore, different manipulations of a tool can determine its use. Manipulating algebraic variables versus describing the qualitative dependencies—these different operations create different products and have different meanings, which are embedded in and contribute to the contexts and culture of their use. Evaluating whether or not the tool was used appropriately will depend on the discipline and the nature of the behaviors and activities that have been and are being endorsed in the community.

As touched upon above, different science disciplines share many features of tool use and the disciplinary boundaries are often blurred when using the same tool. However, the different combinations and weightings of these various features—what tools are used, how they are used, for what purposes, and how to evaluate their use—are centrally grounded in the epistemologies of the disciplines. Moreover, within a large and diverse discipline like biology, there is a great deal of variability among the subdisciplines such as evolutionary biology and molecular biology. Navigating the different ways in which a tool can be used — and figuring out the appropriate use in the many contexts encountered in the undergraduate science curriculum — is no easy task. Therefore, understanding how tools can be used differently to build knowledge will enable instructors to help their students engage in the authentic activities of the different sciences. Furthermore, unpacking the different ways the same tools can be used in physics and in biology provides us a lens that can be used to view the different messages, activities, and problems found throughout the undergraduate science curriculum.

Looking back to the cheetah-pronghorn problem, many physicists would describe solving part (B) of the problem as being physically authentic, reflecting several aspects of the culture of physics. However, there are important elements missing that biologists would look for to consider it biologically authentic. Although there is a backdrop of biology in that living organisms are being considered, this problem does not demand that students engage in using mathematical and physical tools for biological ends. This problem could be situated in a broader discussion on co-evolution in predator-prey relations, thus increasing the biological authenticity of the activity of solving it. In fact, these physics tools and ways of using these tools are important in thinking about natural selection and other biological problems.[4]

---

[3] Of course, it is not as though physicists never attend to higher-order terms and mathematicians never use linearized equations, but that in given contexts and problems, there are occasions in which physicists and mathematicians will attend to different aspects of the same equation.

[4] An updated IPLS problem addressing some aspects of biological authenticity can be found at:

However, this problem was one in a collection of kinematics exam problems in a physics course not explicitly focused on bridging the biology and physics in meaningful ways. These contextual factors all reduce the likelihood that students would engage in meaningful activities connecting the physics to biology.

As the community turns to reform IPLS courses, it becomes imperative to recognize and understand how to support biologically-authentic activities. However, it is not that these activities should replace the physics-authentic activities. Rather, biological authenticity should be used as necessary to help students understand disciplinary differences and build bridges between the various disciplinary practices of biology and physics. In the next sections, we apply our framework on authenticity as a first step in understanding how biology courses ask students to authentically use the tools of physics and how students respond to these uses.

## METHODS: EXAMINING AN INTRODUCTORY BIOLOGY COURSE USING MATH AND PHYSICS

### A. Course description

We examine a reformed introductory biology course, focusing on how the instructors asked students to use mathematics and physics and what messages may be conveyed about their use. The course under study, Principles of Biology III: Organismal Biology (Org Bio), is the third and last course in an introductory biological sciences sequence at University of Maryland. The online catalog describes the course as covering "the diversity, structure, and function of organisms as understood from the perspective of their common physicochemical principles and unique evolutionary histories." The prerequisites are the two preceding courses — or AP credits for these courses — covering topics in cellular and molecular biology and ecological and evolutionary biology.

Org Bio is a relatively new course at the University of Maryland, developed by the faculty to provide a broader perspective on organisms, connecting to and building on the two prior courses. While traditional curricula for this course typically march students through the phyla one-by-one, discussing the characteristic features and functions of each, this course was developed to teach general guiding principles of biology that can be used to understand the differences and commonalities among organisms. The principle most relevant to the use of mathematics and physics is: "Common physical and chemical principles govern all life and nonlife." The instructors of this course weave in mathematics, physics, and chemistry as part of an organizing framework to understand organismal diversity.

In addition to the curricular reforms adopted by the Biological Sciences department, the two courses we examine were also undergoing pedagogical reforms to actively engage students around key concepts, particularly those involving mathematics and physics. The courses met for 50 minutes three times a week. Approximately one-third of the class sessions were devoted to small-group, active-engagement activities. The remaining two-thirds were primarily lecture-based, with a small number of clicker questions supplementing the presentations.

Both of the courses we examined were taught by the same two instructors — a plant developmental biologist and animal evolutionary biologist. The first course was taught in the spring of 2010, with 147 students registered. During the small-group activities, the course was split into two sections of around 75 students each meeting at different times. The second course, taught in fall of 2010, was denoted as an honors course, but retained the same content and structure of the first course. This second course was smaller; only eighty students were registered so that the class was not split for small-group activities.

In Table 1 we show the demographics of both courses, noting that the honors fall course primarily contained freshman students who received AP credit for both pre-requisites, while the spring course was a mix of freshman and sophomores.

|  | spring 2010 | fall 2010 (honors) |
|---|---|---|
| *N* | 147 | 80 |
| freshman | 13% | 71% |
| sophomore | 64% | 21% |
| junior | 18% | 5% |
| senior | 5% | 3% |
| transfer | 26% | 0% |

Table I. Demographics of students enrolled in Org Bio in spring and fall 2010.

---

http://umdberg.pbworks.com/w/page/44332396/The%20cat%20and%20the%20antelope

**Instructor data collection and analysis**

To investigate the ways in which these instructors asked students to use physics and mathematics in this course, we examined the group activity worksheets, homework assignments, and exam problems. For the spring semester, one researcher attended approximately one-third of the classes, while in the fall, one or two researchers attended all class sessions. These observations were used to inform our understanding of the context in which the homework and exam problems were assigned.

To categorize the different uses of mathematics and physics, we first looked at a handful of the homework assignments and exam problems, noting patterns across statements and prompts. In this analysis we looked at the nature of reasoning practices and problem solving that likely would be elicited by the prompts, particularly to construct "correct" responses; we triangulated this with students' written responses and analysis of student interviews. We recognize the methodological limitations of this approach. While we cannot consider the full range of activities in which the students actually engaged as they were solving the problems, we are able to observe student responses in a variety of contexts.

Once we found some consistencies across items, we developed coding categories that reflected the patterns we observed about the use physics and mathematics. With these categories established, we expanded our field to include all the worksheets, homework assignments, and exam problems involving mathematics or physics in the two semesters under study, for a total of seventy-five questions analyzed. These categories were not mutually exclusive and questions could be placed in multiple categories. We refined our categories based on the inclusion of more data and then looked at how these items in each category were distributed with respect to the assignments, the course, and the other categories.

**Student data collection and analysis**

In addition to investigating the ways in which instructors asked students to use mathematics and physics, we also examine students' responses to these tasks. In both semesters, we sent emails recruiting students to participate in one-hour interviews to discuss the course and biology learning in general. In addition, as we were trying to capture a range of students' responses to this course, we purposefully asked students whose participation in class reflected differing opinions toward the course philosophy and inclusion of mathematics and physics. The interviews were loosely-structured and designed to get students to talk qualitatively about their experiences in the course and specific content that had been taught, including reviewing old exam questions. Each student received $10 per interview.

We selected two student interviews to present here as they contained detailed, but differing responses to utility and relatedness of mathematics and physics in biology. Both students, Jenny and Ashlyn, performed well in their respective Org Bio course (B and B+, respectively). Jenny took the honors version in the fall of her sophomore year, while Ashlyn received AP credit and was able to take the course in the spring of her freshman year. Each interview was videotaped and transcribed. We focused on smaller clips in the interview that concerned the inclusion of math and physics. Our qualitative analysis began with discussions of the interview clips in research group meetings and continued with in-depth written descriptions of the discourse that were refined in several iterations among the authors.

**RESULTS: HOW DO THESE INSTRUCTORS ASK STUDENTS TO USE MATHEMATICS AND PHYSICS?**

We developed four categories to organize the various exam and homework questions that incorporated physics and/or mathematics. Each category incorporated different sets of activities with mathematical or physical tools, therefore conveying different messages about how biology students should use mathematics and physics. These categories are based on the epistemological nature of reasoning elicited from students engaged with the problem solving, rather than on what hierarchical level of thought the questions might be designed to elicit. Our focus is on the student behavior in interaction with the learning objectives. We maintain that this range of reasoning has its place in undergraduate science courses. We unpack these categories here, using our lens for understanding authenticity by what tools are used, how they are used, and to what ends.

### Recalling

Several questions on homework and exams asked students to *recall* and state the physical laws or mathematical relationships. These items may have asked students to report the physical laws outright, such as this exam subquestion:

> *Write a concise statement describing the Second Law. If you cite an equation, do not neglect to define the terms and indicate the circumstances for applying the equation, if appropriate. (fall exam 1, question 5A)*

To answer this item, students must recall the Second Law of Thermodynamics and define the relevant terms. While later parts of this question asked students to relate or apply the law in a biological context, this subquestion may convey that part of what biology students need to with physics is *recall* the relevant laws. Other items asked students to describe or draw the graph of the mathematical relationships, without an equation given. For example, on an exam students were asked:

> *In the graph at right, sketch the relationship between volume vs. length (solid line) and surface area vs. length (dashed line). Assume that shape does not change with increasing length. (spring exam 3, question 2A)*

Students must remember that volume grows as length cubed and area as length squared, then graph the respective relationships *or* they must recall the different shapes of the lines representing volume versus length and surface area versus length. Either way, an important component of this problem is simply *recalling* the relationship. Again, this question could send the message that the physical laws or mathematical relationships are things that biology students need to know, without the use of outside materials.

In addition to knowing the specific physical laws and mathematical relationships, several questions asked students to know relevant biological facts that are entwined with or dependent upon physics and mathematics. In these questions, physics and/or mathematics provide the necessary motivation, context, or details for knowing relevant biological concepts. For example, this exam question asked students to state the details of the energy transformations of the sodium potassium exchange pump in cells:

> *Molecules carry out the energy transformations of life. Describe the specific energy transformations that are directly carried out by the Na+/K+ pump. (spring exam 3, question 4A)*

While students were not asked explicitly about the relevant physical laws, physics is woven into the biological details that students needed to write down to answer this question. Students needed to know the specific forms of energy transformation and how that relates to the active transport of ions in and out of the cell. Like the previous examples, this question sends the message that students need to *recall* specific concepts, but in this case those concepts incorporate both physical and biological knowledge.

Questions in the *recalling* category did not ask students to use physical and mathematical tools in meaningful ways, but instead asked them to recall them, characterize them, or know how they combine them with various biological tools. While these questions do not give much insight in tool use, they highlight the importance of what mathematical and physical ideas biology students are told they need to have at their disposal, which may not always be the same tools that are taught in introductory mathematics or physics courses. Furthermore, several questions in this category illustrate how both mathematical and physical tools can be modified by and help modify biological tools, either by merging tools from different disciplines or by specifying the broader context in which other tools are learned and used.

### Synthesizing data

In a few of the in-class small-group activities, students were asked to collect or organize data concerning different physical laws or mathematical relationships. For example, to understand diffusion, students observed a computer simulation portraying random movement of particles. At the start of the simulation, the particles were clumped at the center of the screen; then the particles were allowed to move randomly, with the net result that they tend to spread. Students collected data on the number of particles, their distance from center, and time. The students sent these data to an online spreadsheet in class and then the instructor connected the resulting graphs to the different laws of diffusion. As part of their homework related to this activity, students were asked to synthesize these data again for themselves:

> *On graph paper, plot the data from the simulations attempting to relate the concentration gradient and diffusion rate. Describe the curve seen in your graph, and relate that curve to Fick's First Law. (fall homework 2, question 1)*

Students needed to use the class data to construct a graphical representation, then relate that representation to the mathematical representation of a physical law, Fick's First Law[5], that they were taught in class. To *synthesize* their data, students had to use physical and mathematical tools to develop and compare representations, which differed from just knowing these tools outright in the last category. As in the in-class activity, this homework question may send the message that physics and mathematics can be used to make sense of data that are relevant to biology. This use of the physical and mathematical tools often is found in introductory physics and mathematics courses, pointing to overlap in practices across the science disciplines.

### Calculating

This subset of questions was primarily about use of equations. These items involved the manipulation of different variables and numbers, often to obtain a numerical answer. For example, there were several homework and exam problems asking students to calculate the time it takes molecules to diffuse given distances, such as this one:

> *Flatworms lack circulatory systems so that $O_2$ can only diffuse in their bodies. Assuming a body width of 2.5 cm and a thickness of 1 mm (corresponding to a diffusive distance of 0.5 mm), how much time does it take $O_2$ to diffuse to the center of their bodies?(fall and spring homework 2, question 4A)*

To answer this question, students needed to plug the relevant numbers into an equation discussed in class:

$$t = \frac{x^2}{2D},$$

where *t* is time, *D* is the diffusion coefficient, and *x* is distance. This question mirrors plug-and-chug questions in mathematics and physics courses; it asks students to quantitatively manipulate and insert numbers into the equation to calculate a numerical answer. In thinking about the authentic activities of biology, many questions in this category also show the overlap in tool use across biology, mathematics, and physics.

In addition to the quantitative calculations, we found a few questions that asked students to reason differently with equations:

> *Circle the correct answers to describe the mathematical properties of Fick's First Law.*
>
> *A. If concentration gradient (ΔC/Δx) increases, then the diffusion rate (J) must: increase, decrease or remain the same.*
>
> *B. If concentration gradient (ΔC/Δx) decreases, then the diffusion rate (J) must: increase, decrease, or remain the same.*
>
> *C. A certain concentration gradient for a low molecular weight molecule should result in a: 1) higher, 2) equal, or 3) lower diffusion rate than the same concentration gradient of high molecular weight molecule.*
>
> *(fall homework 2, question 2)*

Instead of calculating a numerical answer, students are asked to qualitatively reason about the equation. Students must manipulate the tool — Fick's First Law — differently than the plug-and-chug of the previous question. They must use the equation to qualitatively determine the proportionalities of different variables.

### Making sense of biological phenomena

Our last category of items was the most prevalent; about half of the homework and exam questions were labeled as using mathematics and physics to *make sense* of biological phenomena. In this organismal biology course, these tools were used to enhance understanding of the structure, function, and evolution of organismal characteristics. For example, the physical and mathematical tools of lever mechanics, plus the mechanical design of the skeletal-muscular system, can be used to understand the advantages of a kangaroo's hop:

> *As kangaroos move from a slow walk to a faster walk their oxygen consumption per time increases. As kangaroos move from walking to hopping, their oxygen consumption actually goes down as they go faster. How can this be? (fall homework 6, question 3)*

---

[5] Fick's First Law, where J is the rate of flux of particles, D is the diffusion coefficient, and ΔC/Δx is the concentration gradient.

To answer this question, students needed to use the physics they had learned in class to examine the in- and out-levers for kangaroo legs, and the ability of those legs to store energy as compressed springs. The conceptual and mathematical tools allowed for greater insight into the structure and function of kangaroo legs and the physical consequences for the breathing rate of these organisms. Unlike the previous categories of questions, in which students were asked to use the mathematics and/or physics to make sense of experimental data or to calculate numerical answers, this question asked students to use physics to better understand organismal biology.

Several *sense-making* questions followed questions from earlier categories, such as *recalling* or *calculating*. For example, after the *calculating* homework question concerning diffusion in flatworms, students had to perform similar calculations for a roundworm and then were posed this question:

> *Using the answers above, describe the constraints that the diffusion of $O_2$ places on the size and shape of these worms, and how they overcome those constraints. (spring and fall homework 2, question 4C)*

Students were asked to use the results of their calculations to better understand how physics relates to the structure and function of worms. To answer this question, they had to apply their earlier calculations on how long it would take oxygen to diffuse to the center of the two different worms and consider the impact of these different times on an organism's ability to survive. The question also asked students to describe what structures have evolved that overcome the constraints that diffusion imposes. Larger organisms would have to endure restrictively long times for oxygen to reach the center of their bodies through simple diffusion and the circulatory systems allow for oxygen to reach to the center more quickly. Therefore, thicker worms, such as the roundworm, have a selective advantage if they have circulatory systems, while thinner worms, such as the flatworm, are not subject to such selective constraints since diffusion allows oxygen to reach the center their bodies in a reasonable length of time. Therefore, in addition to providing greater understanding of structure and function of worms, this question may convey that physics and mathematics can offer insight into organismal diversity.

In both the kangaroo and worm questions, students had to use the tools of mathematics and physics *for different ends* than how these tools are used typically in a physics or mathematics course. Unlike the cheetah-pronghorn problem in our introduction, in these questions physics is used in service of understanding biological phenomena, not just the physical situation. Tying back to disciplinary authenticity, these problems reflect the authentic activities of biology by using mathematical and physical tools for uniquely biological ends.

As a consequence of using these tools for uniquely biological purposes, there are also differences in how students are asked to use the tools in this category. In the following question, students are asked to manipulate the equation differently than when obtaining numerical answers:

> *The Hagen-Poiseuille equation $(V/t) = (\Delta p/R)$ can be used to compare the general features of the circulatory systems of trees vs. mammals. Using the variables in this equation, explain how the circulatory system of mammals is able to generate a higher flow rate than the circulatory systems of trees (fall exam 3, question 2A).*

To answer this question, students needed to dissect the equation to understand the role each variable plays in the rate of flow through a cylindrical pipe. Therefore, the necessary manipulations of this mathematical tool are less like the typical calculating problems, but more like the qualitative reasoning problem described in the previous section. Instead of moving variables around or plugging in numbers, students have to relate the different variables to the biological system and figure out the necessary proportionalities.

Furthermore, the purpose of the tool use can change which features of the tool become salient. The next exam question shows this more explicitly. Students were asked to attend to the variables most relevant to understanding flow in large organisms:

> *Vertebrates (and most other large animals) have separate ventilatory and circulatory systems, each with their own pumps for moving fluids/gases. Gas exchange between the two systems depends on diffusion, as described by the Fick equation: Flow (JA) = -D\*A (Δpp/Δx). For which variable in the Fick Equation is the presence of two pumps (circulatory and ventilatory) most relevant? Why? How does the presence of the two pumps facilitate gas exchange? (fall exam 3, question 1A)*

Instead of thinking of the equation holistically, as would be needed to describe the physical phenomena of diffusion, this question asked students to focus on the relevant variables to understand the biological phenomena. The selective attention to different characteristics of Fick's Law points to a distinctive use of the tool, highlighting the different ways in which these biology students were asked to use equations.

These types of questions represent much of the authentic activities that students were asked to engage in when using mathematical and physical tools in this biology course. While we only examined one introductory biology course that focused on organismal diversity, the differences we found between the biologically-authentic problems and typical physics problems — such as the cheetah-pronghorn problem — emphasize the disciplinary specificity of scientific epistemology, even when using the same tools.

## WHAT DO THESE USES OF MATHEMATICS AND PHYSICS MEAN FOR STUDENTS?

Recognizing that different scientific disciplines may contextualize and interpret activities as authentic in differing ways has significant implications for course development, especially for courses that serve other disciplines effectively (such as a physics or math course for biologists) and courses that effectively integrate disciplines (such as a biology course that incorporates math or physics). Considerations of authenticity are not solely bound to the disciplines, however. In order to develop effectively integrated courses, it is also important to understand how instructors and students interpret and respond to the authenticity of activities [29], [30].

Students often bring into their classes ideas about the nature of the knowledge they are learning and what they have to do to learn it — and these ideas may conflict with the instructor's goals for student learning [13]. Hutchinson [18] offers epistemological authenticity as a useful focus in the classroom for bridging strict disciplinary authenticity with personal authenticity, the sense the student is making of her or his activities independent of the discipline. Among their arguments is that students' framing of their activities influences the ways in which they engage and participate in classroom activities, and attending to how students do this can yield important insights for instruction.

In this section, we begin to address the importance of better understanding and attending to students' perspectives on activities that cross disciplinary boundaries. To do so, we draw on interview data from two students enrolled in the organismal biology course described above, which was designed to include more physics and math. The interviews reveal interesting differences in the ways in which the students responded to, participated in, and framed these expanded and multi-disciplinary practices. Below we consider how these students approached mathematics and physics when asked to use tools in biology-authentic ways, in order to highlight their sense of the relationships among the relevant disciplines.

### Jenny: Understanding physics helps in understanding evolution

We first consider the case of Jenny, a sophomore ecology major, to highlight the possibilities for student epistemologies when using mathematics and physics in biologically-authentic problems. She took Organismal Biology in the first semester of her sophomore year and was interviewed three-quarters of the way through the course. Her interview comments reflect not only her ideas about the interdisciplinary connections and complements in the course, but also how her ideas have changed as a result of engaging with math, physics, and biology in biologically-authentic ways. We chose her interview to present here because she reflects what we believe is representative of what many biology students think about physics and mathematics after traditional science instruction – that biology and physics are separate disciplines with little overlap or relevance to each other. However, the shifts that she notes as a result of her experience in Org Bio show how sophisticated biology students can be in their understanding of how mathematics and physics can be used to better understand biological phenomena.

Throughout the interview, Jenny talked about how she thought her Organismal Biology course was different in how it helped make more explicit the connections among concepts and disciplines, in contrast to her previous biology courses. She highlighted the different principles of the course, describing how they helped organize the biological ideas and provide a framework for understanding evolution. After discussing the importance of phylogenies and the concept of "common ancestors," Jenny volunteered her ideas about mathematics and physics in the course:

> *Jenny: What also made this really different from the AP biology course is that they've used physics and math a lot more. There was that survey that we had to take at the beginning and before we had really started anything in this class and I thought: "Physics and math?! Oh those are completely separate. They don't have anything to do with biology. What are you talking about?" But in this course…I've really been amazed at how many different physics principles and how much more math there is involved than what I thought there was. I knew that…we had used a bit of math in biology [in my AP course] but it was just sort of "oh they threw math in here, what is this?"*

Jenny talked about what her earlier thoughts were about the use of physics and math in biology. She didn't think that biology had "*anything to do with*" physics and math. She recalled how previous uses of math in biology felt disconnected from the rest of biology, just thrown in to the course. These ideas about math and physics in biology are not uncommon. Pre-course survey results suggest that many introductory biology students do not see the connection between biology and math or physics [31]. Furthermore, the treatment of math and physics in biology textbooks often mirrors Jenny's recollection of her AP biology course: the math is put in a box to the side of the main discussion about biology, set apart from the text. Despite her earlier perception about these disciplinary connections, she reported that her experiences in this organismal biology course helped her develop new ideas about how math and physics can be used and what it affords in biology:

> *Jenny: But in this case, they've used physics to explain a lot of the different things so that's been I think the big focus in the class is that there's unity and diversity and you have to figure out how to reconcile those two different things because you have the unity from evolution from the genetics. So you know that the genetics are similar and there's that unity because all the DNA…You can switch DNA in different animals but it's all still the same code. But then when you take different physical principles, that's where you're going to have different…where you're going to have evolutionary changes. Like the dolphins and cows, for example, because there's so many different principles of physics involved with living in those different habitats. Air versus water…so you know that they would have had to have developed different characteristics.*

In contrast to her earlier experiences with math and physics in biology, Jenny described how the instructors in her organismal biology class linked physics with the broader principles of unity and diversity in organismal development. She explained how unity can be understood from thinking about genetics and evolution. Because organisms share so much of the same genomic heritage, there will be similarities among them from their common ancestor. However, to understand the diversity in organisms, Jenny talked about the role of the environment, specifically how different ecosystems can provide the selective pressures resulting in evolutionary changes in organisms. She described how physical principles can shed light on the constraints that different environments pose for organisms, thus providing insight into the diversity in form and function of organismal characteristics. In her account, physics was an integral part of her perspective of organismal biology. Her main point: *Understanding physical principles helps in understanding evolution.*

At this point in the interview, Jenny had described differences in her general attitude toward the role of math and physics in biology, providing an overview of what she thought about biology and physics, without going into much detail. In this next segment of the interview, Jenny was more specific about the physics principles she learned in this course and their link to biology:

> *Interviewer: And so you liked having that addition?*
>
> *Jenny: Yeah I mean I'm not a huge fan of physics but I thought it was sort of "Oh, we don't need to know about that. What's the point of bothering with physics in a biology class?" But using these different things with...let's see we've been talking about diffusion now. The physics and the math behind diffusion...being able to calculate how much time it would take for a molecule to get from...It all depends on the distance. So that helps them to understand...You've got these flatworms that are so flat that they can just diffuse everything through their skin to the center of their body because they're that thin. But when you get animals that are bigger and thicker then you know that diffusion's not gonna work so that's when you know you have to have circulatory systems and different ways of...different gas exchange systems.*
>
> *So we never really incorporated physics in that way...physics and math in that way at all...in my AP biology class because it was just sort of "OK, so these have diffusion but these don't." That might have been mentioned in passing but we never focused on it. It wasn't brought up as much. So we just had to know that the flatworms have diffusion but then roundworms have a circulatory system. Here we actually...by being more hands-on with it and actually going through the equations and figuring out that a molecule has to go this distance and then calculating how much time it takes...Then it clicks, "OK, so you know that this is way too much and it would take days for a molecule of oxygen so they'd be long dead by now so that's why they had to develop a circulatory system.*
>
> *So that really helped to connect it a lot too because in the AP class it was just "OK these just need diffusion they're fine but then these ones they have to have a circulatory system." Here it's connected with the different physics principles and actually calculating through and actually figuring out...That's why there's a circulatory system. That sortofa thing. Instead of just learning "OK these do these don't."*

Jenny again referred back to her pre-class attitudes about physics, citing her ideas that it was superfluous for understanding biology. She then began to recount how principles of diffusion relate to organismal development. She described that the physics and math "behind diffusion" helps her understand the constraints that organism face in ion transport and gas exchange. She highlighted the relationship between diffusion time and distance, claiming that "being able to calculate" helps her to understand the development of flatworms and other animals. Different organisms evolve in different ways to deal with the physical constraints of their environment; understanding the physical constraint of diffusion helped her explain the selection pressure favoring the origins of circulatory and gas exchange systems.

Beyond helping her understand the biological content, Jenny also elaborated how learning physics principles helped her change her approach to learning biology; she explicitly referred to the role physics had on her epistemology of biology. Instead of learning that some animals "have diffusion" and some don't, the physics principles helped her understand *why* different organisms have different methods for circulating oxygen throughout the body. In particular, Jenny stated that the process of using the equations to calculate the time it takes for molecules to diffuse different distances helped her piece together a biological explanation for the organismal diversity. For Jenny, biology became less about memorizing lists and more about constructing a story for understanding the evolution of different characteristics and mathematics and physics help her build and tell that story.

While not all of the students in Org Bio were as sophisticated or articulate in their ideas about the role math and physics has in biology, Jenny exemplifies how bridging the disciplines can help students develop their understanding of what it means to think like a scientist. In Jenny's case, her ideas about what it means to learn biology have, in her own estimation, developed through using physics and mathematics in biologically-authentic ways.

While Jenny expressed a very favorable response to the use of mathematics and physics in Org Bio, not all of the students were as positive. To provide an opposing perspective, we chose the case of Ashlyn, a freshman biology major who received AP credit for both of the prerequisite courses. Ashlyn was interviewed about halfway in the spring-semester course, after a small-group activity and series of lectures covering diffusion. We present her interview comments here as a contrast to Jenny's, but also to highlight some of the epistemological challenges that students may face as they are asked use mathematics and physics tools in biology.

### Ashlyn: Equations are for mathematics and physics, not biology

In this interview, Ashlyn revealed that she had not taken a biology course in the past year, but she said she liked the subject. In particular, she talked about how she chose the biology major because she thought it was more relevant to the real world. Ashlyn contrasted her ideas about biology with those of chemistry, talking about how she appreciated that she could "*perceive*" biology, as compared to chemistry, which she said was "*under the microscope*."

Throughout the interview, she mentioned that she felt she was struggling in Org Bio, but thought the course was interesting and that she was "*learning all this new stuff, plus… learning to see it in a different light.*" After talking more about her experiences and study habits in the course, Ashlyn was asked about the recent use of equations in class. She responded that she had "*blocked out*" the equations so far. She elaborated on why:

> *Ashlyn: I don't like to think of biology in terms of numbers and variables. I feel like that's what physics and calculus is for. So, I mean, come time for the exam, obviously I'm gonna look at those equations and figure them out and memorize them, but I just really don't like them.*
>
> *Interviewer: Ok. So you've blocked them out and you don't like them, keep going.*
>
> *Ashlyn: I understand, like, what they're used for, what they do, but the actual placement--*
>
> *Interviewer: And that is?*
>
> *Ashlyn: -- like for diffusion and gas exchange and stuff, but I don't remember precisely what the variables and what the equation is.*

In the first part of the quote, Ashlyn voiced her distaste for the use of equations in biology. She drew disciplinary boundaries around what tools are used — equations, with numbers and variables, are the tools of physics and mathematics, not biology. Ashlyn then talked about how she used these tools in this biology course: she had to figure out and memorize the equations for the exam.[6] She elaborated that she knew the purposes of the equation, but struggled with knowing exact details of the tool itself. In this segment, the use of equations that Ashlyn focused on was *recalling* specific aspects of the tool, such as the definitions of the variables. She expanded further on the details that she felt she needed to know:

> *Interviewer: Is it a matter of memorizing the variables that's a problem for you?*
>
> *Ashlyn: It's memorizing how they fit together. If you give me, like, for example, like, the diffusion equation on the last exam, if you*

---

[6] Her use of "*obviously*" is interesting here. Although she doesn't like the equations or think that they are part of biology, she states that she will (obviously) engage with them for the exam. It also could be that what is obvious is what she needs to do with the equations to be successful: recognize and memorize them. She also could be using "obviously" as a way to interact with the interviewer, suggesting that she thought her expectations would be shared between her and the interviewer.

*gave me the units, I could figure it out for the most part, but the equation with the letters that stand for numbers, sometimes I can't remember which letters stand for what, and where they go, but I do remember, like, what goes where. I know that distance goes on top, and the--*

*Interviewer: You want to draw it?*

*Ashlyn: Hold on. It was x squared over 2d, and distance goes on top and that's the diffusion constant, and I remember that because I just looked it at before coming here, but if I hadn't done that, then I would just know that the distance that it goes travels on top, and I would not necessarily remember the letters that go in that place, so I guess I have a more, like, broad and less detail-oriented knowledge of the equations.*

*Interviewer: So do you think the equations are necessary to understanding how diffusion works?*

*Ashlyn: Kind of, I mean, it's basically a way to put it, put the concept into words. I think that's what the only function of the equations are. It's just to help you write it down. If you understand that the distance that it goes is on-- like, if you just look at it in terms of units even, it would be easier for me to remember than just to write down a couple of letters.*

Ashlyn elaborated on how it is necessary to memorize the placement of the different letters, as well as what those letters represent. Interestingly, she claimed to be able to figure out the equation if she were given units. In doing so, she contrasted two ways of reproducing this tool: (1) remembering "what stands for what" and "what goes where" and (2) reasoning it out using the units or the referents themselves, such as distance. She talked about how she felt she had a broad understanding of the equations, but not the knowledge of the details, which is what she thought she was expected to recall for the exams.

Ashlyn also stated that the only function of equations is: to "*put the concept into words… help you write it down.*" In contrast to Jenny, who spoke about how the equations and physical principles could be used to help her better understand biology, Ashlyn talked about equations only as referents or memory devices for the physical concepts. She claimed that "*that's… the only function of the equations.*" She continued on to talk about diffusion specifically:

*Ashlyn: And like I said, I think that biology is just-- it's supposed to be tangible, perceivable, and to put that in terms of letters and variables is just very unappealing to me, because like I said, I think of it as it would happen in real life, like if you had a thick membrane and you try to put something through it, the thicker it is, obviously the slower it's gonna go through. But if you want me to think of it as this is x and that's d and then this is t, I can't do it. Like, it's just very unappealing to me.*

In this last quote, Ashlyn reiterated her ideas about what biology is "supposed to be," contrasting those ideas with her ideas about equations. She used the example of diffusion to make her point. In "real life" —which tied into her ideas about biology being tangible and perceivable — it is easy to understand that the thicker a membrane is, the longer it's going to take to go through. She claimed that using letters and numbers to represent this idea is unappealing; it strips away the real world from understanding biology. Interestingly, the functional relationship — that the time-to-diffuse is proportional to the square of the distance — was absent in her explanation.

When talking about this same example, Jenny expressed very different views about the usefulness of the diffusion equations. First, she talked about different uses of the equations than Ashlyn, including using them to calculate the diffusion time for different organisms. Jenny claimed that this calculation helped her better understand and make connections in biology. The quantitative aspects of mathematics and physics were powerful for Jenny, while Ashlyn did not find that they added to her understanding.

In this segment of her interview, there are several epistemological challenges that Ashlyn expressed that were consequential for her response to the use of equations in this course. First, Ashlyn solely focused on equations as something to know or to help recall a given concept. She did not bring up or discuss other uses of this tool in Org Bio. Second, she expressed that she saw disciplinary boundaries around the different tools used in the sciences -- equations are for physicists and mathematicians, not biologists. Finally, she only focused on the qualitative aspects of this mathematical tool, not the functional relationships which can help further explain the physical constraints of diffusion in organisms. While later in the interview Ashlyn was more positive about the use of mathematics in biology — which highlights the context-dependence of student ideas — this segment emphasized how challenging it can be for students to use mathematics and physics tools across disciplinary lines.

## DISCUSSION

As undergraduate biology and pre-medical education undergoes transformations to meet the challenges set forth by policy-makers and post-graduate schools, IPLS instructors and physics education researchers must keep up with the evolving needs of life-science students. In addition to examining the content needs of these students, the physics education community must also attend to the epistemological challenges these students face as they use the same concepts and tools across scientific disciplines. Understanding the authentic practices of biology, physics, and mathematics will enable instructors to make explicit the disciplinary differences and help students navigate and make connections between these in and out of the classroom.

While we situate authenticity in the practices of experts, we are not advocating that students participate in the overt activities of physicists and biologists. Instead, we heed the words of Dewey in recognizing that classroom activities must blend the authentic aspects of the disciplines with the experiences and developmental needs of students [31]. We are "concerned not with the subject matter as such, but with the subject matter as a related factor in a total and growing experience." (p. 30). In this paper, we offer disciplinary authenticity as a lens to understand the subject matter, recognizing that this is not the only focus needed in examining the educational experiences of students. We offer an analytical approach for unpacking disciplinary authenticity, looking at the different ways in which disciplines use tools for different scientific ends. We then use this framework to analyze an organismal biology course, documenting the authentic ways that a biology course asked students to use mathematics and physics tools. This course used these tools in many ways, but the most dominant use was in service of understanding biological phenomena and processes. Equations were used to help understand form and function of different biological systems and characteristics, while students were asked to use physics topics to describe how physics constrains the evolution of organismal design, as well as how physics is exploited in organismal function. The physics and mathematics were part of a larger endeavor to understand the complex world of organismal diversity.

In contrast, the opening problem with the cheetah and pronghorn were in service of understanding the relationships between accelerating objects and those at constant velocity — valuable aspects for understanding the physics of a situation, but not necessarily the biology. While the opening paragraph nods to the broader biological context of the predator-prey co-evolution, the questions themselves do not demand that students engage in the practices of understanding the advantages and constraints that kinematics offer for organisms. This problem is likely constructive for helping students learn how to "think like a physicist," for the reasons described in the earlier analysis, but does not make the connection to understanding how to apply these tools to "think like a biologist." While it might be too much to ask of a physics course that it teach students how to "think like a biologist", the multi-disciplinary authenticity perspective does imply that it might be appropriate to create physics courses that help biology students perceive the authentic biological value of learning physics.

The results from the organismal biology course provide insight into how to make these connections. For example, after a series of calculating exercises using Fick's First Law, the students were asked to apply these results to discuss how they related to the evolutionary constraints on the size and shape of worms. Furthermore, we consistently found that the *synthesizing* and *calculating* tasks were presented as a part of a larger endeavor to understand organismal design. The broader situational and course contexts can help shape the individual activities in which students engage while completing a specific task or answering a question [32], [33]. Our analysis of the cheetah-pronghorn problem in isolation suggested that it would not elicit biologically-authentic practices from students; taking into account the broader context of the exam and the course further suggests that the student activities will not be meaningfully connected to the biology. We argue in this paper that helping students cross disciplinary epistemological boundaries will require keeping the physics-authentic activities, but tying them more authentically to the practices of biologists. Furthermore, these connections are important at multiple levels: in the design of individual tasks and in considering the broader homework/exam and course contexts.

Alongside the differences we found in disciplinary practices between this biology course and the typical physics course, we also found differences in how the students responded to use of mathematics and physics tools. Extensive work has been conducted in physics education to document and describe the expectations, epistemologies, and attitudes of introductory physics students. In particular, many students in introductory physics courses express that they see equations as disconnected from physics concepts [8], [9], [12]. Interestingly, the two organismal biology students presented here expressed very different views about the relationship between math and physics concepts. When Ashlyn expressed negative views about the role of physics and mathematics in biology, it was not because she did not recognize the deep conceptual underpinnings of the equations. In fact, on several occasions she explicitly linked the equation as a way to express concepts. Instead, she stated that she did not see how the equation added value beyond the representing conceptual relationships — which is an expectation not documented in the physics education literature. This discrepancy is not surprising; research has shown that student epistemologies can be context- and discipline-specific [34], [35]. Our results from these students' responses to these biologically-authentic activities highlight the need for research not just on the variability of students' epistemological expressions, but also on the context in which they are expressed.

Physics educators and education researchers have new challenges to meet in order to respond to the calls for biology education reform. IPLS courses are being updated to include more of the physics and mathematics tools that are relevant for life-science majors. In this paper, we advocate for and take first steps in thinking about this reform more broadly, by exploring and making connections to the authentic activities of biologists and by examining student responses to these new environments.

## ACKNOWLEDGMENTS

We gratefully acknowledge conversations with the members of the University of Maryland Physics and Biology Education Research Groups and with numerous faculty in the College of Life Sciences and Education. This material is based upon work supported by the US National Science Foundation under Awards No. DUE 03-41447. Any opinions, findings, and conclusions or recommendations expressed in this publication are those of the author(s) and do not necessarily reflect the views of the National Science Foundation